# Multi-layered Social Networks

**Piotr Bródka**, Przemysław Kazienko

Wrocław University of Technology, Institute of Informatics

Wrocław

Poland

**piotr.brodka@pwr.wroc.pl**, kazienko@pwr.wroc.pl

## Synonyms

Multi-layered Social Networks,
Layered social network,
Multi-relational social network,
Multidimensional social network,
Multiplex social network

## Glossary

*SN* – social network

*SSN* - single-layered social network

*MSN* – multi-layered social network

## Definition

A social network $(SN)^1$ is defined as a tuple $<V,E>$, where:

$V$ – is a not-empty set of actors representing social entities: humans, organizations, departments etc., called also vertices, nodes, vertexes, members, cases, agents, instances or points;

$E$ – is a set of directed edges (relations between actors called also arcs, connections, or ties) where a single edge is represented by a tuple $<x,y>$, $x,y \in V$, $x \neq y$ and for two edges $<x,y>$ and $<x',y'>$ if $<x,y> \neq <x',y'>$ and $x=x'$ then $y \neq y'$. Note that $<x,y> \neq <y,x>$ because we consider here directed social networks.

Since social *networks* usually represent one kind of relationships they are also called *single-layered social network SSN* [Magnani 11].

A multi-layered social network *MSN* is a network extended to multiple edges between pairs of nodes/actors. It is defined as a tuple $<V, E, L>$ where:

$V$ – is a not-empty set of actors (social entities);

---

[1] In this essay *a social network* (*SN*) is also called *a single-layered social network* (*SSN*) to distinguish it from *a multi-layered social network* (*MSN*), see below

*E* – is a set of tuples <*x,y,l*>, *x,y*∈*V*, *l*∈*L*, *x*≠*y* and for any two tuples <*x,y,l*>, <*x',y',l'*>∈*E* if <*x,y,l*>≠<*x',y',l'*>, *x*=*x'* and *y*=*y'* then *l*≠*l'*;

*L* – is a fixed set of distinct layers (types of relationships).

# Introduction

It is quite obvious that in the real world, more than one kind of relationship can exist between two actors (e.g. family, friendship and work ties) and that those ties can be so intertwined that it is impossible to analyse them separately [Fienberg 85], [Minor 83], [Szell 10]. Social networks with more than one type of relation are not a completely new concept [Wasserman 94] but they were analysed mainly at the small scale, e.g. in [McPherson 01], [Padgett 93], and [Entwisle 07]. Just like in the case of regular single-layered social network there is no widely accepted definition or even common name. At the beginning such networks have been called multiplex network [Haythornthwaite 99], [Monge 03]. The term is derived from communications theory which defines multiplex as combining multiple signals into one in such way that it is possible to separate them if needed [Hamill 06]. Recently, the area of multi-layered social network has started attracting more and more attention in research conducted within different domains [Kazienko 11a], [Szell 10], [Rodriguez 07], [Rodriguez 09], and the meaning of multiplex network has expanded and covers not only social relationships but any kind of connection, e.g. based on geography, occupation, kinship, hobbies, etc. [Abraham 12].

Social networks emerging from different types of social media or social networking sites are good examples of multi-relational networks. A main reason for that is the fact that these systems offer large and diverse datasets including information about people profiles and their various activities that can be analysed in depth. Since this data reflects users' behaviours in the virtual world, the extracted social networks are called online social networks [Garton 97], web-based social networks [Golbeck 06], or computer-supported social networks [Wellman 96].

Bibliographic data, blogs, photos sharing systems like Flickr, e-mail systems, telecommunication data, social services like Twitter or Facebook, video sharing systems like YouTube, Wikipedia and many more are the examples of services providing data sources which may be used by many researches to analyse underlying social networks. However, this vast amount of data and especially its multi-relational character simultaneously present new research challenges related to processing problems of this data [Domingos 03]. Although most of the existing methods work properly for single-layered networks, there is a lack of well-established tools for multi-layered network analysis. Development of new measures is very important from the perspective of further advances in the web science as the multi-relational networks can be found almost everywhere. They are more expressive in terms of the semantic information and give opportunity to analyse different types of human relationships [Rodriguez 09].

# Key Points

Each layer in the multi-layered social network *MSN* corresponds to one type of relationships between people. Different relationships can result from the character of connections, types of communication channel, or types of various collaborative activities that humans (e.g. users of various IT services) can perform within a given system or in a given environment. The examples of different relationships can be: friendship, family or work. Different communication channels that result in different types of connections are: email exchange, VoIP calls, instant messenger chats, etc. The separate relationship types can also be defined based on users' common activities within complex services like in the photo publishing service. Since users may publish their photos, comment pictures provided by others, add some photos to their favourites, then each such activity can reflect various kinds of relationships: author-commentator, commentator-commentator, author-favourite, etc. Additionally relationships can also possess semantic meaning as for example publishing and commenting photos is a much more proactive action (more semantic) than just adding photos to favourites often utilized due to acquaintance with the author, see [Kazienko 2011a]. Another example where information about user activities has a clear semantic meaning can be an internet forum where people, who are very active and post a lot of queries can be perceived as new to a field. On the other hand, people who comment a lot but do not post any queries can be seen as experts in a forum domain.

Actors $V$ and all edges $E_l \subseteq E$ from only one layer $l \in L$ correspond to a simple, single-layered social network $SSN= <V, E_l, \{l\}>$. In general, a multi-layered social network $MSN=<V,E,L>$ may be represented by a multigraph, where multiple relations are represented by multiedge [Newman 10]. Hence, all the structural measures presented below can also be applied to other kinds of complex networks that are described by means of multigraphs.

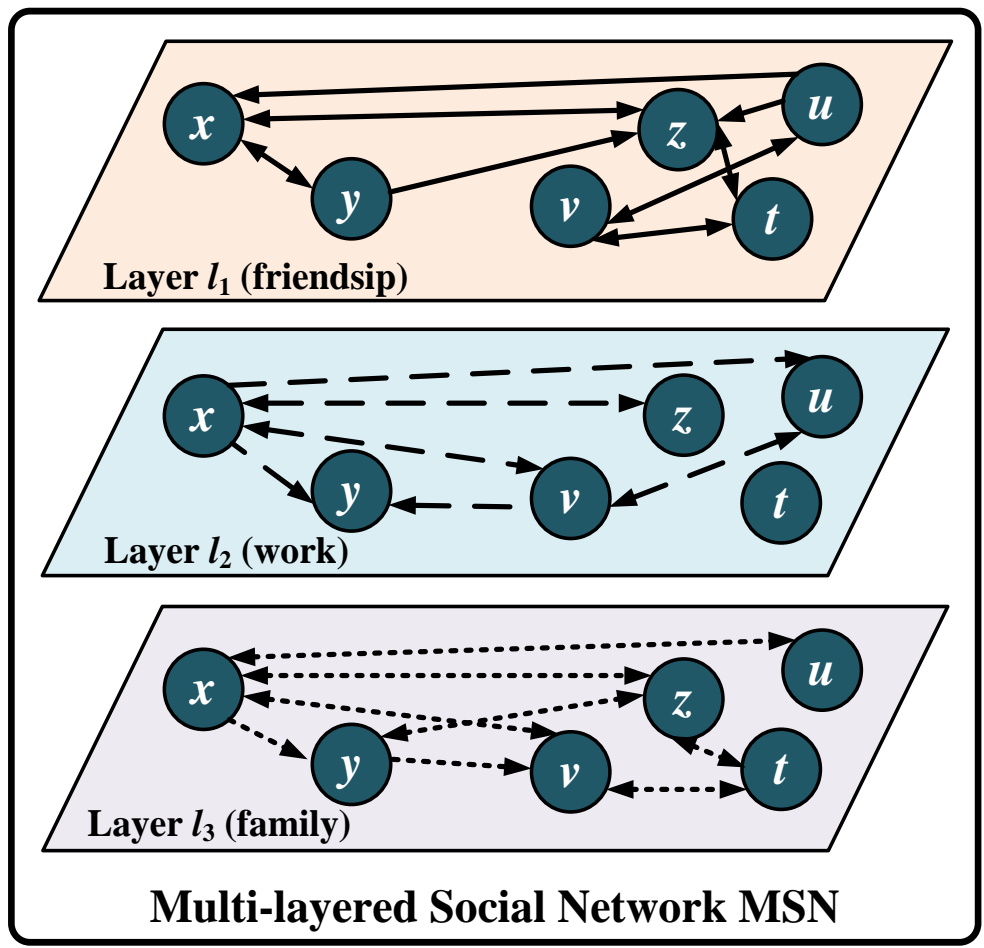

Figure 1. An example of the multi-layered social network *MSN*.

In Figure 1, the example of three-layered social network is presented. The set of actors consists of {*t, u, v, x, y, z*} so there are six actors in the network that can be connected with each other on three layers: $l_1$ (friendship), $l_2$ (work) and $l_3$ (family). On the layer $l_1$, eight relationships (tuples) between actors: <$x,y,l_1$>, <$y,x,l_1$>, <$x,z,l_1$>, <$z,x,l_1$>, <$y,z,l_1$>, <$u,z,l_1$>, <$u,v,l_1$>, <$v,u,l_1$> can be distinguished, 6 edges on layer $l_2$ and 7 on layer $l_3$. The multi-layered social network from Fig. 1 represents a quite dense family network ($l_3$) that is simultaneously less intensive collaboration network ($l_2$) and a bit more crowded friendship structure ($l_1$).

## Historical Background.

Social networks with more than one kind of relationship between the same actors may take many different names. The most common one is *Multi-Layered* (or just *Layered*) *social networks* [Bródka 12], [Geffre 09], [Hamill 06], [Kennedy 09], [Magnani 11], [Schneider 11] but also *Multi-relational social networks* [Szell 10], *Multi-dimensional social networks* [Kazienko 11b], *Multidimensional Temporal social network* [Kazienko 11c], or *Multivariate social networks* [Szell 10] are in use.

Additionally, the researchers in the field of multi-layered networks also try to develop new models of networks that capture not only the multi-layered profile of social data. Authors in [Cantador 06] proposed a multi-layered semantic social network model that enables to investigate human interests in more details than when they are analysed all together. Wong at al. in [Wong-Jiru 07] propose multi-layered model, where only one layer describes relation between people, the other layers are processes, applications, systems and a physical network.

Despite the fact that many researchers investigate multi-layered social networks there is surprisingly few definitions or descriptions going beyond the statement "*It is the network where two nodes are connected by more than one connection, relation, tie*". Newman [Newman 10] defines such a network as a multigraph with multiedges between nodes. A network is then represented by the adjacency matrix, whose each cell $A_{ij}$ (a multiedge) actually only preserves a simple number of edges between node *i* and *j*. The second possible representation is an adjacency list where a multiedge is represented by multiple identical entries in the list of neighbours, e.g. an adjacency list (*x,v,y,y,y,z,z*) means that node *x* has one edge to node *v*, three edges to y, and two to *z*. The main problem with Newman's definition is that he has never assigned any labels on those edges so in this model the information about which edge represents what component single-layered network is lost.

A different approach is presented by Magnani and Rossi in [Magnani 11]. They describes two concepts: (1) *Pillar Multi-Network* and (2) *Multi Layer Network* (*ML-Model*). The first concept defines social network with multiple relation as a set of single-layered networks {<$V_1,E_1$>, <$V_2,E_2$>, …, <$V_k,E_k$>} and some mapping relation by means which a node in one network is mapped to another node in the second network. It means that one node from one single-layered network $SN_1$ can reflect only one node in another single-layered social network $SN_2$. This case is typical for most web-based services, e.g. a Facebook account may correspond to the Twitter account. The second concept is almost the same but the mapping function is slightly different, i.e. many nodes (users) from one single-layered social network $SN_1$ can match to a single node in another single-layered social network $SN_2$. For example, if one single-layered social network $SN_1$ represents relations between co-workers from one company and another network $SN_2$ characterizes relations between whole departments in the same company, then many employees working in a given department *D* (many nodes in $SN_1$) are mapped to this single department *D* (one node) in the second network $SN_2$.

Kazienko at al. in [Kazienko 11b] [Kazienko 11c] present yet another model of multidimensional temporal social network, which considers three distinct dimensions of social networks: layer, time and group dimension. All the dimensions share the same set of nodes that corresponds to social entities: single humans or groups of people. A layer dimension describes all kinds of relationships between users of the system; a time dimension reflects the dynamics of the social network and a group dimension focuses on interactions within separated social communities (groups). At the intersection of all this dimensions is a small social network, which contains only one kind of interactions (one layer) for a particular group in a given time snapshot. This concept allows to analyse systems, where people are linked by many different relationship types (layers in the layer dimension) like in complex social networking sites, e.g. Facebook. It means, people may be connected as friends, via common groups, "like it", etc. It can also refer complex relationships within regular companies: department colleagues, best friends, colleagues from the company trip, members of particular project team, etc. Multidimensionality provides an opportunity to analyse each layer separately and at the same time investigate different aggregations over instances of the layer dimension. For example, let us consider a network composed of six layers, three from the real word: family ties, work colleagues and gym friends and three from the virtual world, i.e. friends from Facebook and fiends from the MMORPG game as well as friends from some discussion forum. Now, one has many different possibilities for studies on such a network, for example: (i) to analyse each layer separately, (ii) to aggregate layers from the real world and compare them to the virtual world layers aggregation, and finally, (iii) to aggregate all layers together. Additionally, a time dimension provides possibility to investigate the network evolution and its dynamics. For example, the analysis (i) how users neighbourhoods change when one of the neighbours leaves the network and how it affects the network in longer term, (ii) how roles of group leaders (e.g. project team leaders) change over time – are overtaken by different people, or (iii) how changes on one layer affect the other layers. Finally, the group dimension allows studying groups existing within the social network. Using multidimensionality, not only the usual social groups can be analysed (friend family, school, work, etc.) but also groups created upon various member features like gender, age, location, etc. Moreover, the model allows to compare the results of different community extraction methods, e.g. by means of graph-based social community extraction or typical data mining clustering. To conclude, the multidimensional social network enables to analyse all three dimensions at the same time, e.g. how interaction on different five layers of two social groups changes over three selected periods.

Rodriguez in [Rodriguez 07] defines a multi-relational social network as a tuple $G = (N, E, W)$, where *N* is the set of nodes in the network, *E* is a set of directed edges, *W* is the set of weights associated with each edge of the network $|W|=|E|$. However, two years later in [Rodriguez 09] he defined the same network as $M = (V, E)$, where *V* is the set of vertices in the network, $E = \{E_1, E_2, ..., E_m\}$ is a family of edge sets in the network, and any $E_k \subseteq (V \times V): 1 \leq k \leq m$. Each edge set in *E* has a different semantic interpretation.

Therefore, it is crucial to define one general concept of the multi-layered social network. The idea further presented is very similar to *Pillar Multi-Network* presented in [Magnani 11], but it was introduced one year earlier in [Kazienko 10]. The main difference is that instead of mapping function, in this model the set of nodes is unified and fixed, i.e. it is common for all layers. The appropriate definition was presented above in the Definition Section.

# Measures in Multi-layered Social Network

[…] To see the rest of the paper please contact the authors or Springer

# Acknowledgements [optional].

The work was partially supported by the Polish Ministry of Science and Higher Education, the research project 2010-13.

# References.